\documentclass[conference]{IEEEtran}
\IEEEoverridecommandlockouts
\usepackage{cite}
\usepackage{amsmath,amssymb,amsfonts}
\usepackage{algorithmic}
\usepackage{graphicx}
\usepackage{textcomp}
\usepackage{xcolor}
\usepackage{url}
\usepackage[caption=false,font=footnotesize]{subfig} 
\def\BibTeX{{\rm B\kern-.05em{\sc i\kern-.025em b}\kern-.08em
    T\kern-.1667em\lower.7ex\hbox{E}\kern-.125emX}}

\newcommand{\polarisfootnote}{\footnote{\label{fn:polaris}\url{https://www.alcf.anl.gov/polaris}}}
\newcommand{\polarisref}{\textsuperscript{\ref{fn:polaris}}}
\begin{document}

\title{Toward Distributed 3D Gaussian Splatting for High-Resolution Isosurface Visualization}


\author{\IEEEauthorblockN{Mengjiao Han}
\IEEEauthorblockA{\textit{Argonne National Laboratory} \\
USA \\
hanm@anl.gov}
\and 
\IEEEauthorblockN{Andres Sewell}
\IEEEauthorblockA{\textit{Utah State University} \\
USA \\
a02024444@usu.edu}
\and
\IEEEauthorblockN{Joseph Insley}
\IEEEauthorblockA{\textit{Argonne National Laboratory} \\
USA \\
insley@anl.gov}
\and
\IEEEauthorblockN{Janet Knowles}
\IEEEauthorblockA{\textit{Argonne National Laboratory} \\
USA \\
jknowles@anl.gov}
\and
\IEEEauthorblockN{Victor A. Mateevitsi}
\IEEEauthorblockA{\textit{Argonne National Laboratory } \\
\textit{University of Illinois Chicago} \\
USA \\
vmateevitsi@anl.gov}
\and
\IEEEauthorblockN{Michael E. Papka}
\IEEEauthorblockA{\textit{Argonne National Laboratory} \\
\textit{University of Illinois Chicago} \\
USA \\
papka@anl.gov}
\and
\IEEEauthorblockN{Steve Petruzza}
\IEEEauthorblockA{\textit{Utah State University} \\
USA \\
steve.petruzza@usu.edu}
\and
\IEEEauthorblockN{Silvio Rizzi}
\IEEEauthorblockA{\textit{Argonne National Laboratory} \\
USA \\
srizzi@anl.gov}
}

\maketitle

\begin{abstract}

We present a multi-GPU extension of the 3D Gaussian Splatting (3D-GS) pipeline for scientific visualization. Building on previous work that demonstrated high-fidelity isosurface reconstruction using Gaussian primitives, we incorporate a multi-GPU training backend adapted from Grendel-GS to enable scalable processing of large datasets. By distributing optimization across GPUs, our method improves training throughput and supports high-resolution reconstructions that exceed single-GPU capacity. 
In our experiments, the system achieves a 5.6× speedup on the Kingsnake dataset (4M Gaussians) using four GPUs compared to a single-GPU baseline, and successfully trains the Miranda dataset (18M Gaussians) that is an infeasible task on a single A100 GPU. 
This work lays the groundwork for integrating 3D-GS into HPC-based scientific workflows, enabling real-time post hoc and in situ visualization of complex simulations.

\end{abstract}


\section{Introduction}
Recent advances in 3D-GS~\cite{kerbl3Dgaussians} enable photorealistic, real-time rendering of complex 3D scenes by optimizing anisotropic Gaussian primitives from view-dependent images. 
%
Recent works have started applying 3D-GS for scientific data visualization~\cite{ai2025nli4volvis,tang2025ivr,sewell2024high,yao2025volseggs}.
However, their pipeline was limited to a single-GPU setting, restricting its scalability and performance, particularly for large and complex datasets typically generated on high-performance computing (HPC) systems.

In this work, we present a multi-GPU extension of the 3D-GS pipeline tailored for scientific visualization. Building on the work by Sewell et al.~\cite{sewell2024high}, our approach adapts the parallel training strategy from Grendel-GS~\cite{zhao2024scaling}, a distributed framework originally designed for large-scale scene reconstruction. 
%
Our contributions are as follows: (1) We present multi-GPU 3D-GS pipeline for large-scale scientific data on HPC systems. (2) We demonstrate the effectiveness of our parallel implementation with benchmarks across multiple image resolutions and varying number of GPUs. (3) We provide an initial step toward scalable, real-time scientific data reconstruction using 3D-GS on HPC platforms, establishing a foundation for future extensions to multi-node and in situ applications.

\section{Background and Related Work}
3D-GS’s speed and efficiency have led to growing interest in scientific visualization.
Sewell et al.\cite{sewell2024high} first applied 3D-GS to scientific data, reconstructing high-fidility isosurfaces visualizations. After that, prior works extended 3D-GS to editable volume rendering~\cite{tang2025ivr}, LLM-based interaction~\cite{ai2025nli4volvis}, and dynamic scene segmentation~\cite{yao2025volseggs}.
Despite these advances, none of the existing studies have addressed the use of multiple GPUs or distributed HPC environments for 3D-GS-based scientific visualization. Given the scale and complexity of scientific datasets, which are often generated on HPC systems using multiple nodes, there is a pressing need for scalable visualization solutions. In this study, we present an initial investigation into distributed 3D-GS training and rendering across multiple GPUs, aiming to facilitate efficient 3D reconstruction of large-scale scientific data.

\section{Method Overview}
Following \cite{sewell2024high}, we extract isosurface point clouds directly from scientific volume data using ParaView. 
%
These point clouds are used as initializations for Gaussian primitives, bypassing the need for image-based reconstruction. 
A set of synthetic camera views is generated in a structured orbit, and the splatting model is trained to match these views. 
%
We extend this pipeline using Grendel-GS’s~\cite{zhao2024scaling} distributed training design, where each GPU holds a shard of the global point cloud and Gaussian parameters. 
The training data is distributed across workers, and gradients are synchronized via a fused all-reduce scheme. 
By adapting the distributed training of 3D Gaussian splatting, we can scale our method to multiple GPUs and make it happen on larger scale of scientific data. 

\section{Results}

Two datasets were used in our study: Kingsnake (110.3 MB, about 4M points) and Miranda (491 MB, about 18.18M points). Training was performed on a single Polaris HPC node at Argonne, equipped with four NVIDIA A100 GPUs\polarisfootnote.
While prior work by Sewell et al.~\cite{sewell2024high} used ~250 images per dataset, we increased this to 448 to better support the complexity of our data. 
To assess multi-GPU performance, we trained with image resolutions of 512\texttimes512, 1024\texttimes1024, and 2048\texttimes2048, and benchmarked training time and image quality using 1, 2, and 4 GPUs (Table \ref{tab:kingsnake} and \ref{tab:miranda}).

%
%


\begin{figure}[htbp]
\vspace{-2em}
  \centering
  \subfloat[Ground truth]{%
    \includegraphics[width=0.475\linewidth]{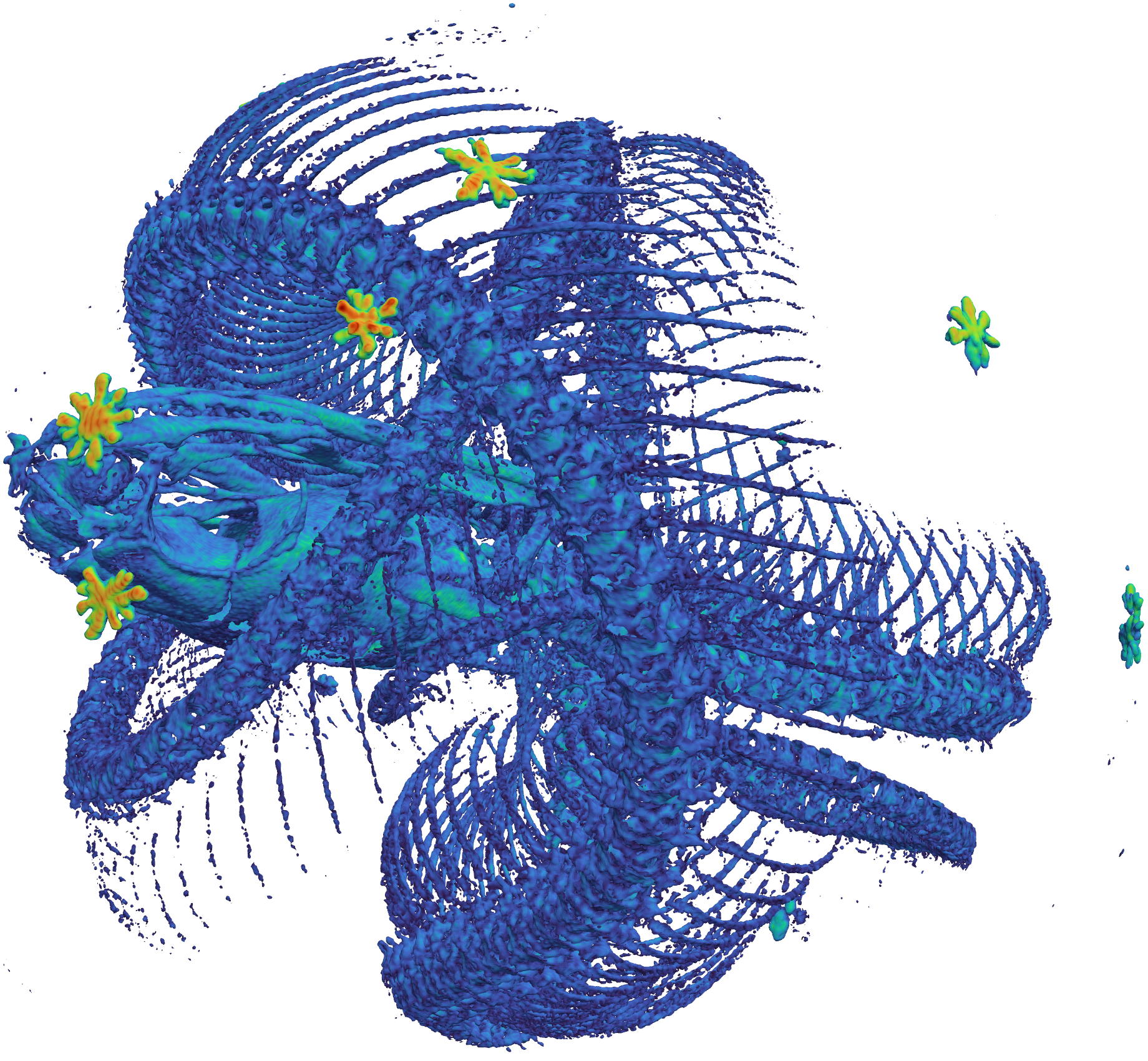}
    \label{fig:subfigA}
  }
  \hfill
  \subfloat[3D-GS rendered]{%
    \includegraphics[width=0.475\linewidth]{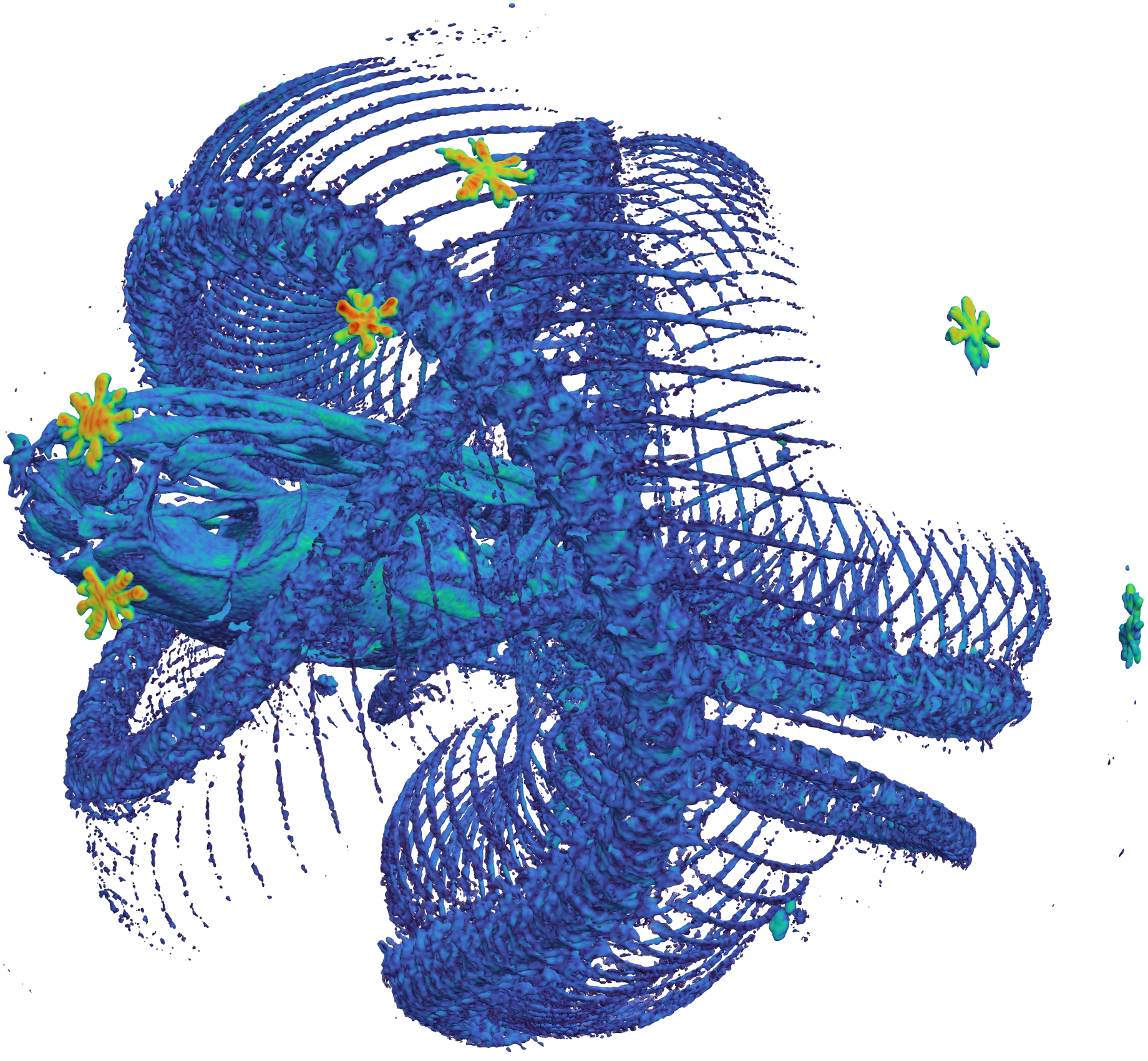}
    \label{fig:subfigB}
  }
  \caption{Ground truth isosurface versus 3D-GS rendering of the Kingsnake dataset at 2048×2048 resolution. Training was performed with 4 A100 GPUs on Polaris at Argonne\polarisref. The result achieves an average PSNR of 29.32, SSIM of 0.97, and LPIPS of 0.03.\vspace{-0.5em}
  }
  \vspace{-1em}
  \label{fig:vis-kingsnake}
\end{figure}

Figure~\ref{fig:vis-kingsnake} compares the ground truth isosurface visualization with the 3D-GS rendered result for the Kingsnake dataset. 
3D-GS achieves high reconstruction quality, with average metrics of PSNR 29.32, SSIM 0.97, and LPIPS 0.03 for Kingsnake, and PSNR 36.37, SSIM 0.9905, and LPIPS 0.011 for Miranda. A detailed comparison of image quality is provided in Tables \ref{tab:kingsnake} and \ref{tab:miranda}.
Tables~\ref{tab:training_time} present the training times for the Kingsnake and Miranda datasets across different image resolutions and GPU counts. The results clearly demonstrate that increasing the number of GPUs significantly reduces training time, particularly for high-resolution inputs such as 2048\texttimes2048 (e.g., a 5.6× speedup for Kingsnake at 2048×2048 using 4 GPUs vs. 1). 
Additionally, as Zhao et al.~\cite{zhao2024scaling} note, a single A100 GPU supports up to approximately 11.2M Gaussians. Consequently, larger datasets like Miranda, around 18M Gaussians, exceed this memory limit and require multi-GPU training. This highlights the necessity and practical value of employing distributed 3D-GS for large-scale scientific visualization.


%


\begin{table}[]
\caption{Training time (minutes) for Kingsnake and Miranda datasets at different image resolutions and GPU counts. 'X' indicates failure on a single A100 GPU due to memory limits.}
\label{tab:training_time}
\resizebox{\columnwidth}{!}{%
\begin{tabular}{ccccccc}
\hline
  & \multicolumn{3}{c}{Kingsnake ($\sim$4M Gaussians)} & \multicolumn{3}{c}{Miranda ($\sim$18M Gaussians)} \\ \hline
\#GPUs &
  512 &
  1024 &
  2048 &
  512 &
  1024 &
  2048 \\ \hline
1 & 12.60           & 18.60           & 48.00          & X               & X              & X              \\
2 & 11.01           & 10.48           & 15.46          & 20.37           & 21.88          & 50.10          \\
4 & 6.07            & 5.97            & 8.50           & 11.60           & 12.20          & 16.84          \\ \hline
\end{tabular}%
}
\vspace{-2em}
\end{table}


\begin{table}[]
\caption{PSNR ($\uparrow$), SSIM ($\uparrow$), and LPIPS ($\downarrow$) for the Kingsnake dataset across different image resolutions and GPU counts.}
\label{tab:kingsnake}
\resizebox{\columnwidth}{!}{%
\begin{tabular}{cccccccccc}
\hline
       & \multicolumn{3}{c}{512} & \multicolumn{3}{c}{1024} & \multicolumn{3}{c}{2048} \\ \hline
\#GPUs & PSNR   & SSIM  & LPIPS  & PSNR    & SSIM  & LPIPS  & PSNR    & SSIM  & LPIPS  \\ \hline
1      & 25.52  & 0.95  & 0.056  & 26.90   & 0.96  & 0.056  & 25.12   & 0.93  & 0.089  \\
2      & 25.87  & 0.96  & 0.046  & 28.63   & 0.97  & 0.035  & 29.33   & 0.97  & 0.030  \\
4      & 25.87  & 0.96  & 0.046  & 25.03   & 0.93  & 0.067  & 29.32   & 0.97  & 0.030  \\ \hline
\end{tabular}%
}
\vspace{-2em}
\end{table}

\begin{table}[]
\caption{PSNR ($\uparrow$), SSIM ($\uparrow$), and LPIPS ($\downarrow$) for the Miranda dataset across different image resolutions and GPU counts.}
\label{tab:miranda}
\resizebox{\columnwidth}{!}{%
\begin{tabular}{cccccccccc}
\hline
       & \multicolumn{3}{c}{512} & \multicolumn{3}{c}{1024} & \multicolumn{3}{c}{2048} \\ \hline
\#GPUs & PSNR   & SSIM  & LPIPS  & PSNR    & SSIM  & LPIPS  & PSNR    & SSIM  & LPIPS  \\ \hline
2      & 31.62  & 0.99  & 0.014  & 34.21   & 0.99  & 0.010  & 36.30   & 0.99  & 0.011  \\
4      & 31.63  & 0.99  & 0.014  & 34.22   & 0.99  & 0.010  & 36.37   & 0.99  & 0.011  \\ \hline
\end{tabular}%
}
\vspace{-2em}
\end{table}


\section{Conclusion and Future Work}

We presented a multi-GPU extension of the 3D-GS pipeline for scientific data visualization. Leveraging the Grendel-GS parallel training strategy, our implementation enables scalable optimization of Gaussian primitives across multiple GPUs, efficiently handling large volumetric datasets that exceed single-GPU memory limits. For instance, we achieve a 5.6× speedup on the Kingsnake dataset with 4 GPUs and successfully train the 18M-Gaussian Miranda dataset, which is infeasible on a single A100 GPU.
Our results demonstrate that multi-GPUs 3D-GS significantly improves training throughput and scalability without compromising reconstruction quality. This work marks the first step toward adapting 3D-GS for real-time, large-scale scientific visualization on HPC platforms.

Looking ahead, we plan to extend our framework to support multi-node deployments across distributed HPC environments, enabling reconstruction of even larger datasets. Additionally, we aim to explore integration with uncertainty quantification methods to capture reconstruction confidence and investigate the feasibility of in situ rendering to further reduce storage and I/O overhead in scientific workflows.

\section*{Acknowledgment}

This research used resources of the Argonne Leadership Computing Facility, a U.S. Department of Energy (DOE) Office of Science user facility at Argonne National Laboratory and is based on research supported by the U.S. DOE Office of Science-Advanced Scientific Computing Research Program, under Contract No. DE-AC02-06CH11357.

\bibliographystyle{IEEEtran}
\bibliography{reference}


\end{document}